\documentclass[10pt]{article}
\usepackage{express}
\usepackage{wasysym}
\usepackage{color}
\begin{document}
	
	\newcommand{\SiOx}{SiO$_2$}
	\newcommand{\TaOx}{Ta$_2$O$_5$}
	\newcommand{\um}{\textmu m}
	\newcommand{\Po}{\ensuremath{P_0}}
	
	\title{UV superconducting nanowire single-photon detectors with high efficiency, low noise, and 4~K operating temperature}
	
	\author{E.~E.~Wollman,\authormark{1,*} V.~B.~Verma,\authormark{2} A.~D.~Beyer,\authormark{1} R.~M.~Briggs,\authormark{1} F.~Marsili,\authormark{1} J.~P.~Allmaras,\authormark{1} A.~E.~Lita,\authormark{2} R.~P.~Mirin,\authormark{2} S.~W.~Nam,\authormark{2} and M.~D.~Shaw\authormark{1}}
	
	\address{\authormark{1}Jet Propulsion Laboratory, California Institute of Technology, 4800 Oak Grove Dr., Pasadena, California 91109, USA\\
		\authormark{2}National Institute of Standards and Technology, 325 Broadway, Boulder, CO 80305, USA}
	
	\email{\authormark{*}Emma.E.Wollman@jpl.nasa.gov} %% email address is required
	
	
	
	\begin{abstract}
		For photon-counting applications at ultraviolet wavelengths, there are currently no detectors that combine high efficiency (> 50$\%$), sub-nanosecond timing resolution, and sub-Hz dark count rates. Superconducting nanowire single-photon detectors (SNSPDs) have seen success over the past decade for photon-counting applications in the near-infrared, but little work has been done to optimize SNSPDs for wavelengths below 400~nm. Here, we describe the design, fabrication, and characterization of UV SNSPDs operating at wavelengths between 250 and 370~nm. The detectors have active areas up to 56~\um~in diameter, 70 - 80$\%$ efficiency, timing resolution down to 60~ps FWHM, blindness to visible and infrared photons, and dark count rates of $\sim$ 0.25~counts/hr for a 56~\um~diameter pixel. By using the amorphous superconductor MoSi, these UV SNSPDs are also able to operate at temperatures up to 4.2~K. These performance metrics make UV SNSPDs ideal for applications in trapped-ion quantum information processing, lidar studies of the upper atmosphere, UV fluorescent-lifetime imaging microscopy, and photon-starved UV astronomy.
	\end{abstract}

	%%%%%%%%%%%%%%%%%%%%%%%%%%  body  %%%%%%%%%%%%%%%%%%%%%%%%%%
	\section{Introduction}
	
	For time-correlated single-photon counting at near-IR wavelengths, superconducting nanowire single-photon detectors (SNSPDs) have surpassed all other detector technologies. At 1550~nm, SNSPDs combine high detection efficiencies ($> 90\%$ \cite{Marsili:2013}), low noise (intrinsic dark count rates $< 0.1$~counts per second \cite{Marsili:2013}), high timing resolution (jitter as low as 15~ps FWHM \cite{Zadeh:2016}), and high dynamic range (maximum count rates of over 100~MHz for a four-pixel array \cite{Rosenberg:2013}). SNSPDs operating at visible and IR wavelengths have been used for fundamental tests of quantum mechanics \cite{Shalm:2015}, laser ranging \cite{Warburton:2007}, CMOS fault detection \cite{Somani:2001}, fluorescence microscopy \cite{Yamashita:2014}, and optical communication \cite{Boroson:2014}. One practical downside of SNSPDs is the additional complexity required to cool the devices down to cryogenic temperatures. In particular, efficiencies > $80\%$ have only been reported at temperatures below 3~K \cite{Marsili:2013, Verma:2015, Zhang:2016, Zadeh:2016}.
	
	While much of the focus of SNSPD development has been at IR wavelengths, the fundamental SNSPD detection mechanism should also work well at much shorter wavelengths. When an incident photon is absorbed in the superconducting nanowire, it forms a local region of suppressed superconductivity in the wire. If the nanowire is current-biased sufficiently close to its critical current density, this region can cause a section of the nanowire to go normal, which in turn causes the bias current to be redirected into the readout circuit \cite{Goltsman:2001}. Shorter-wavelength photons deposit more energy in the wire, allowing for the use of thicker nanowires with higher critical temperatures, higher critical currents, and lower dark count rates. With the development of optical stacks to enhance absorption into the nanowire layer, similar to those used at near-IR wavelengths, UV SNSPDs have the potential to improve on the noise performance and operating temperature of near-IR SNSPDs while maintaining high efficiencies. 
	
	At UV wavelengths (< 400~nm), there are few detectors capable of counting single photons with high efficiency, low noise, and precise timing. Photomultiplier tubes (PMTs), for example, have sub-nanosecond time resolution, but are less than $50\%$ efficient. Delta-doped electron-multiplying charge coupled devices (EMCCDs) have been optimized for $80\%$ efficiency in the UV \cite{Nikzad:2016}, but can only operate at the high frame rates required for fast photon counting with considerable increases in noise. Replacing PMTs with SNSPDs optimized for UV operation will lead to improvements in trapped-ion quantum information processing, fluorescence lifetime imaging, and UV lidar.
	
	Recently, SNSPDs have been developed for ion-trapping experiments at 315~nm\cite{Slichter:2017}. Here, we present MoSi detectors optimized for 370~nm operating above 4~K with $>80\%$ detector efficiency and system dark count rates below 0.01~counts/s. We discuss the effects of detector size and operating temperature on efficiency, optical bandwidth, and timing performance.
	
	\begin{figure}[b!]
	\centering\includegraphics[]{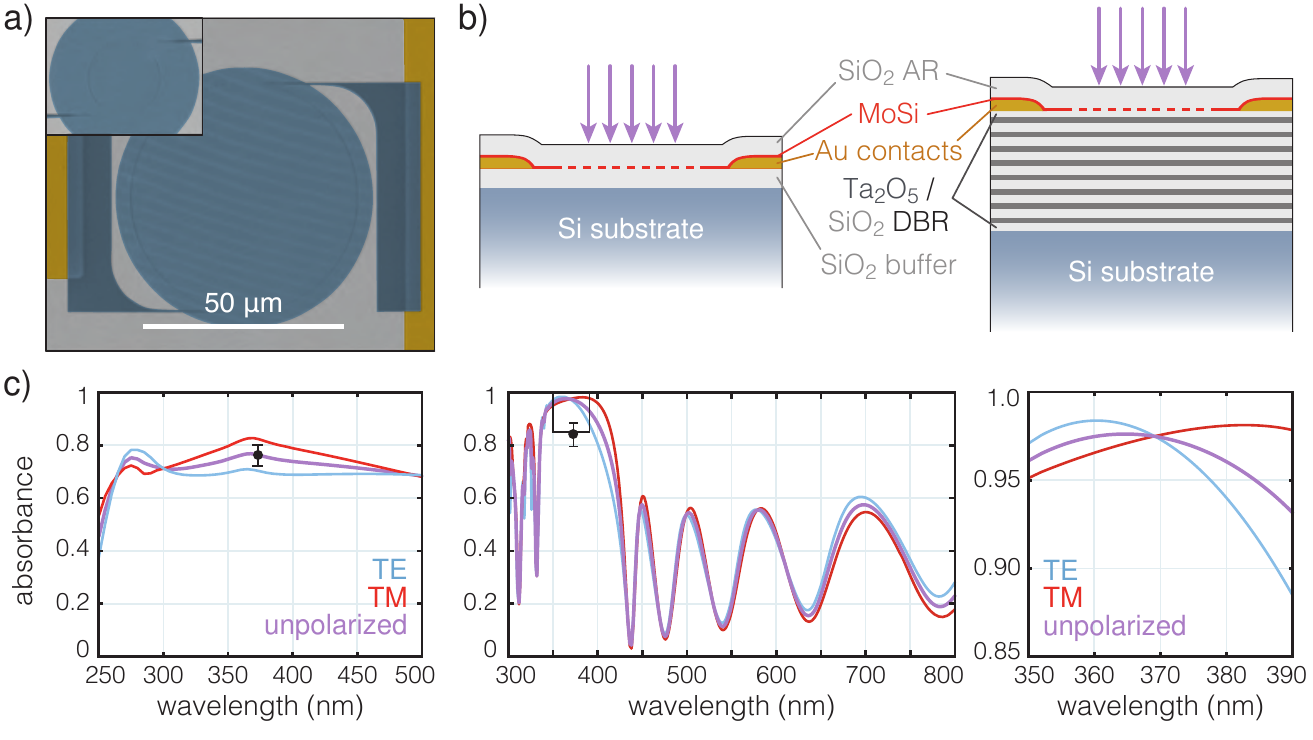}
	\caption{a) False-color SEM images of nanowire patterns for large and small (inset) active area SNSPDs. Both images are to scale. b) Optical stack cross-section for broadband (left) and narrowband (right) devices. c) RCWA simulation of absorption by the nanowire layer for TE-polarized (blue), TM-polarized (red), and unpolarized (purple) light. For TE-polarized light, the electric field is oriented parallel to the wires. The left panel shows the predicted absorption for the broadband stack from (b), while the center and right panels show the predicted absorption for the narrowband stack. In the narrowband design, near the target wavelength of 370 nm, absorption for both polarizations is equal. Black data points indicate measured device detection efficiency.}
	\label{design}
\end{figure}
	
	\section{Design}
	Moving to shorter wavelengths creates both challenges and simplifications for SNSPD design and fabrication. Materials used in optical stacks at near-IR wavelengths no longer have the desired performance below 400~nm, and nanowire dimensions on the scale of the wavelength can lead to diffraction effects. Due to the larger energy per photon, however, it is possible to saturate the internal device efficiency for wires with larger cross-sections. Larger wires are easier to fabricate with high yield, provide higher signal-to-noise at a given temperature, and allow for operation at higher temperatures. We found that a wire cross-section of 10~nm $\times$ 110~nm maximized the operating temperature while still saturating the internal device efficiency at 370~nm. The nanowire pitch was kept below 200~nm to avoid diffraction effects. We made devices with either a 56~\um~or a 16~\um~diameter active area for coupling to either 50~\um~or 10~\um~core multimode fiber (Fig.~\ref{design}a).
	
	To enhance absorption by the nanowire layer, we used a single-layer anti-reflection (AR) coating on top of the device layer and a dielectric mirror below it (Fig.~\ref{design}b). We fabricated devices with two different optical stack designs: the first used the silicon wafer as a back-reflector to produce moderately-high broadband efficiency from 275~nm to 400~nm, making it suitable for detecting fluorescence from Mg$^+$, Be$^+$, or Yb$^+$ ions; the second used a distributed Bragg reflector (DBR) to produce higher efficiency over a narrow band centered at the Yb$^+$ resonant wavelength of 370~nm. The total absorption by the MoSi layer for unpolarized light is predicted to be $\sim75\%$ for the broadband devices and 96\% for the narrowband ones using rigorous coupled-wave analysis (RCWA) \cite{Zhang:2010} (Fig.~\ref{design}c). At 370~nm, the efficiency of the narrowband devices is predicted to be polarization independent. For either stack design, it is possible to optimize for higher efficiencies at a fixed polarization if the application allows for polarization control of the incoming light. Based on measurements of the optical properties of MoSi films deposited at JPL, it should also be possible to design optical stacks with >~$70\%$ broadband absorption in the device layer using aluminum back-mirrors and MgF AR coatings at wavelengths down to at least 200~nm. The optical properties of sputtered Mo$_{0.75}$Si$_{0.25}$ films have not yet been studied at wavelengths below 200~nm.
	
	Throughout the following sections, we discuss results from devices with different combinations of optical stacks and active areas. For simplicity, we will refer to these detectors as D1, D2, and D3, with the properties of each described in Table \ref{tab}.
	
	\begin{table}[t!]
		\renewcommand{\arraystretch}{1.25}
		\centering
		\begin{tabular}{c c c c c}
			\hline
			Device & Diameter (\um) & Optical stack design & SDE at 373~nm & DDE at 373 nm \\ \hline
			D1     & 56 & narrowband & $76 \pm 4\%$ & $84 \pm 5\%$\\
			D2     & 56 & broadband & $69 \pm 4\%$ & $76 \pm 4\%$ \\
			D3     & 16 & narrowband & $69 \pm 4\%$ & $80 \pm 4\%$\\
			\hline
		\end{tabular}
		\caption{Definition of device parameters for detectors presented in this article.}
		\label{tab}
	\end{table}
	
	\section{Detector Efficiency}
	To characterize devices at 370~nm, two sources were used. The first was an incoherent LED with center wavelength $\sim 375$~nm. The second was a PicoQuant LDH-P-C-375 pulsed laser diode with a center wavelength of $373$~nm. The light from either source was collimated, directed through two attenuator wheels consisting of various neutral density (ND) filters, and then coupled into a multimode fiber patch cable via a reflective coupler. The patch cable could be connected either to a powermeter for calibration, or to a fiber at the top of the cryostat leading to the SNSPD. 
	
	We define the system detection efficiency (SDE) as the ratio of the count rate of the detector to the estimated photon arrival rate at the top of the fridge --- that is, the probability that a photon leaving the end of the patch cable at the top of the cryostat will produce a measureable click. To measure the system detection efficiency, we first calibrated the ND filters by turning up the laser or LED power and measuring the optical power in the fiber with and without a given filter in place. We then turned down the source power and measured the power in the fiber with no filters in place (\Po). We then attenuated the beam to the desired photon flux with a combination of filters from the two wheels. Using the measured \Po, our filter calibrations, and a correction for the back-reflected light at the end of the fiber during the power measurement, we calculated the rate of photons entering the cryostat. When the pulsed laser was used to measure the SDE, the mean number of photons per pulse was kept below 0.002 so that any statistical deviations from a continuous wave source were less than $0.1\%$. The error of our efficiency measurement was largely set by the accuracy of the powermeter, which is estimated to be $\sim 5\%$. To test the accuracy and linearity of our attenuator calibrations, we produced the same predicted photon flux using different combinations of \Po~and attenuators, and we measured the count rate from the SNSPD. We found that the measured count rate was independent of choice of attenuators within a $1.7\%$ standard deviation. The total SDE error is thus $5.3 \%$ of the SDE.
	
	\begin{figure}[t!]
		\centering\includegraphics[]{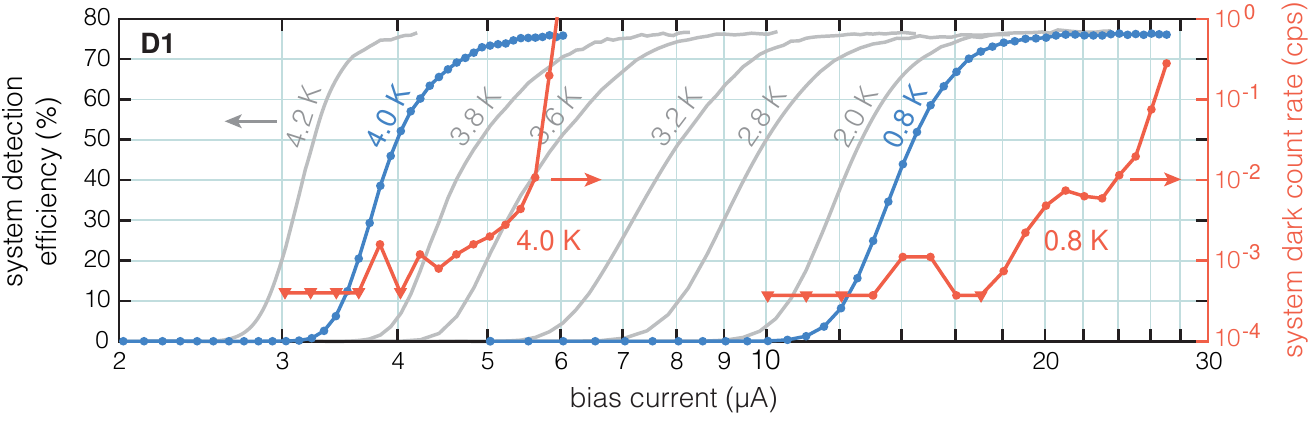}
		\caption{Left axis, gray and blue data: system detection efficiency vs. bias current at different temperatures for D1. Error bars of $\pm 4\%$ are not shown. For all temperatures, the SDE reached $76\%$. Taking into account losses in the fiber, this corresponds to an efficiency of $84\%$ at the device. Right axis, red data: system dark count rate vs. bias current at 800~mK and at 4.0~K (corresponding SDE shown in blue). Triangles indicate no detected counts in the measurement window. At 800~mK, dark count rates were $\sim 10^{-3}$ counts/s (cps) at the onset of the plateau. At 4.0~K, dark count rates were less than $10^{-2}$ counts/s at the onset of the plateau.}
		\label{tempdep}
	\end{figure}
	
	Figure \ref{tempdep} shows the measured SDE at 370~nm for a large active area, narrowband device (D1) as a function of both bias current and temperature. For both narrowband and broadband devices, we observed a plateau in efficiency vs. bias current at temperatures up to 4.2~K, indicating saturation of the internal efficiency. For narrowband D1, the SDE reached $76\pm 4\%$, and for broadband D2, the maximum SDE was $69\pm 4\%$. Possible sources of loss include scattering at the connection between the patch cable and the cryostat fiber, loss in the fiber itself, scattering at the fiber-device interface, coupling inefficiencies to the nanowire layer in the optical stack, and internal inefficiencies in the nanowire detection mechanism. 
	
	As fiber loss is not negligible at ultraviolet wavelengths, we also calibrated out the loss in the fiber to find the device detection efficiency (DDE). The device detection efficiency is the probability that a photon leaving the end of the cryogenic fiber will produce a measurable click, and includes losses due to absorption or reflection in the optical stack and internal device inefficiencies. Depending on the application, splicing to the cryostat fiber, using shorter lengths of fiber, AR-coating the end of the fiber, or free-space coupling to the device may be possible, and so the DDE provides an estimate of the optimal device performance. The loss in the cryostat fibers was calibrated at room temperature by measuring the power sent into the cryostat fiber with a powermeter at the top of the cryostat, and comparing it to the power measured by a powermeter at the position of the device. We confirmed that the fiber transmission does not change measurably at low temperatures by coupling two fibers together at the cold stage of the cryostat and measuring the transmission through the fibers at both room temperature and at the base temperature of the cryostat. Using this calibration, we found that the fiber transmission was $94.2\pm0.1\%$. This gave an estimated DDE of $84 \pm 5\%$ for D1 and $76 \pm 4 \%$ for D2 at 370~nm.
	
	For many applications, it is helpful to have detectors that operate well above the base temperature of a pulse-tube cryocooler. As cryogenic ion traps, for example \cite{Vittorini:2013}, tend to have large heat loads, operation at or above 4~K is ideal. As seen in Figure~\ref{tempdep}, while the switching current of the device decreased at higher temperatures, the detection efficiency was still observed to reach the same value at 4.2~K as it did at the cryostat base temperature.
	
	\begin{figure}[t!]
		\centering\includegraphics[]{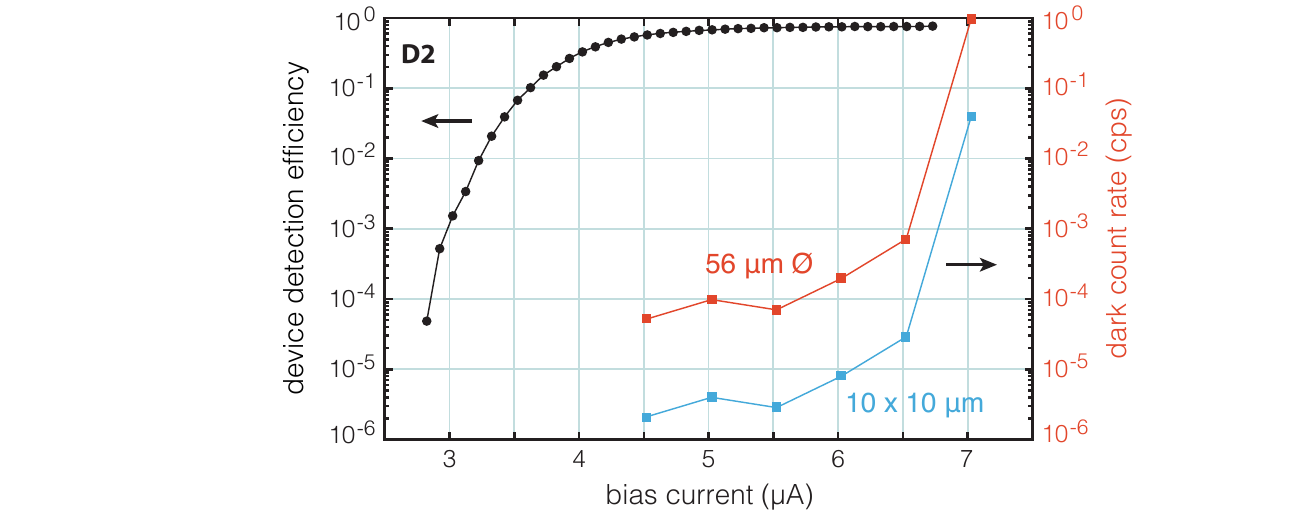}
		\caption{Red squares: dark count rate for a large active area, broadband device (D2) disconnected from the optical fiber. The device detection efficiency (black circles) is shown for reference. At the onset of the efficiency plateau ($\sim 5$ \textmu A), the dark count rate is approximately $10^{-4}$ counts/s. Also shown is the same data scaled for the hypothetical area of a $10 \times 10$ \um~pixel (blue squares).}
		\label{dcr}
	\end{figure}
	
	\section{Dark count rates}
	To measure system dark count rates, we replicated a typical laboratory test environment: the device was connected to the room-temperature optical set-up via an optical fiber, the laser was turned off, the filter wheel blanked, and the room lights turned off. At 800~mK, dark counts for a large active area, narrowband device (D1) were at the 0.001~counts/s level at the onset of the efficiency plateau (Fig.~\ref{tempdep}). At 4~K, dark counts were higher, but still below 0.01~counts/s at the onset of the plateau.
	
	With the device linked to the room temperature environment via the optical fiber, dark count rates were likely dominated by stray photons entering the fiber. For applications such as cryogenic ion trapping where the light source is in the same cryostat as the detector, or in free-space coupled applications with low backgrounds, it is also helpful to measure the dark count rate when the device is not connected to room temperature via a fiber. Figure \ref{dcr}~shows the dark count rate for a large active area, broadband device (D2) measured at 4~K without a fiber connection. At the onset of the efficiency plateau, the dark count rate was at the $10^{-4}$ counts/s level. It is also worth noting that the dark count rate is expected to scale with device area. For a 10~\um~$\times 10$~\um~pixel suitable for coupling to a single-mode UV fiber or for use in an imaging or spectroscopy array, the dark count rate would thus be $\sim 3\times10^{-6}$ counts/s or $0.01$~counts/hr, comparable with microchannel plate detectors.
	
	\section{Mid-UV response and out-of-band rejection}
	To characterize devices at shorter wavelengths, we used UV LEDs at 250~nm, 285~nm, and 315~nm. The LEDs were filtered to reject out-of-band wavelengths and coupled into a 50~\um~core multimode fiber attached to the detector.
	
\begin{figure}[t!]
		\centering\includegraphics[]{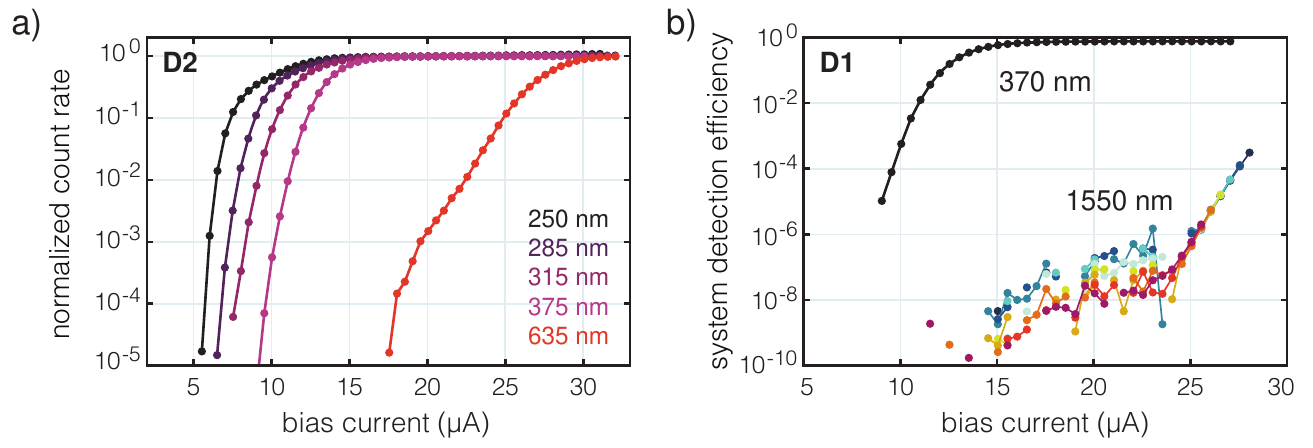}
		\caption{a) Normalized count rate vs. bias for a large active area, broadband device (D2) measured at 800~mK. Data were taken with LEDs at 250~nm, 285~nm, and 315~nm, with a laser diode at 373~nm, and with a laser at 635~nm. b) System detection efficiency vs. bias for a large active area, narrowband device (D1) measured at 800~mK. Black points were taken at 373~nm, and colored points were taken with a 1550~nm laser at different powers. Below $\sim 24$~\textmu A, the count rate at 1550~nm is dominated by noise.}
		\label{wldep}
	\end{figure}

	Figure \ref{wldep}a shows the count rate vs. bias at 800~mK for a large active area, broadband UV SNSPD (D2) under illumination at several UV wavelengths. The count rate was fitted with an error function and normalized by its asymptotic value. The device was single-photon sensitive and produced a plateau in the count rate vs. bias, indicating a saturation of the internal efficiency, at all wavelengths from 250~nm to 373~nm. For shorter wavelengths, this saturation happens at lower currents. As the dark count rate decreases with decreasing current, operating at shorter wavelengths allows for lower dark counts without sacrificing efficiency. As discussed above, the device detection efficiency for this device was estimated to be $76\pm 4\%$ at 373~nm, and similar efficiencies are expected down to 275~nm (see Fig.\ref{design}c). Also shown is the response to a 635~nm laser. At a bias current of 17~\textmu A, the internal efficiency at 635~nm is a factor of $10^4$ lower than at 250~nm - 375~nm. This blindness to longer wavelengths for a device with saturated internal efficiency at 375~nm suggests that SNSPDs designed for far-UV wavelengths should be solar-blind --- a useful feature in detectors for UV astronomy.
	
	In Figure~\ref{wldep}b, the system detection efficiency vs. bias current is shown for a narrowband device (D1) measured with both a 373~nm laser and a 1550~nm laser. At 1550~nm, the detection efficiency is suppressed by at least a factor of $10^7$ relative to the efficiency at 373~nm at the onset of the plateau. This level of suppression at near-IR wavelengths makes UV SNSPDs insensitive to dark counts from blackbody radiation entering the optical fiber at room temperature, leading to the low system dark count rates discussed in the previous section.
	
	\section{Timing}
	The 50~\um~core multimode fiber used in the above measurements has relatively low loss and is easier to couple to than single-mode UV fiber. In order to cover such a large active area, however, the nanowire is required to be 14 mm long. The resulting kinetic inductance of almost 10~\textmu H leads to relatively long electrical rise and decay times for the SNSPD pulse, and thus to poor timing characteristics compared with a typical near-IR SNSPD.
	
	We also fabricated 16~\um~diameter narrowband devices (D3) suitable for coupling to a 10~\um~core 0.1~NA optical fiber. Due to the increased fiber losses, the measured SDE for these devices was close to $70\%$, with the estimated DDE still over $80\%$. These devices were $\sim 10\times$ shorter than the large active area devices, with a corresponding improvement in electrical time constants. Below $\sim 2.5$~K, the 16~\um~$\diameter$ devices exhibited evidence of latching. At higher temperatures, some combination of lower bias currents, higher kinetic inductance, and increased thermal conductance enhanced the device's thermal recovery time relative to its electrical recovery time and prevented latching.
	
	\subsection{Dead time}
	The dead time of the detector was characterized by illuminating the device with the 375~nm UV LED and measuring the time between consecutive pulses. At 800~mK, a histogram of these times reveals a window of $\sim 90$~ns ($\sim 11$~ns) for the 56~\um~$\diameter$ (16~\um~$\diameter$) device where no inter-photon arrival times were measured. At this temperature, the pulse fall time was $\sim 250$~ns ($\sim 25$~ns). The dead time affects the maximum count rate that can be obtained before the detection efficiency begins to decrease.
	
	\begin{figure}[ht!]
		\centering\includegraphics[]{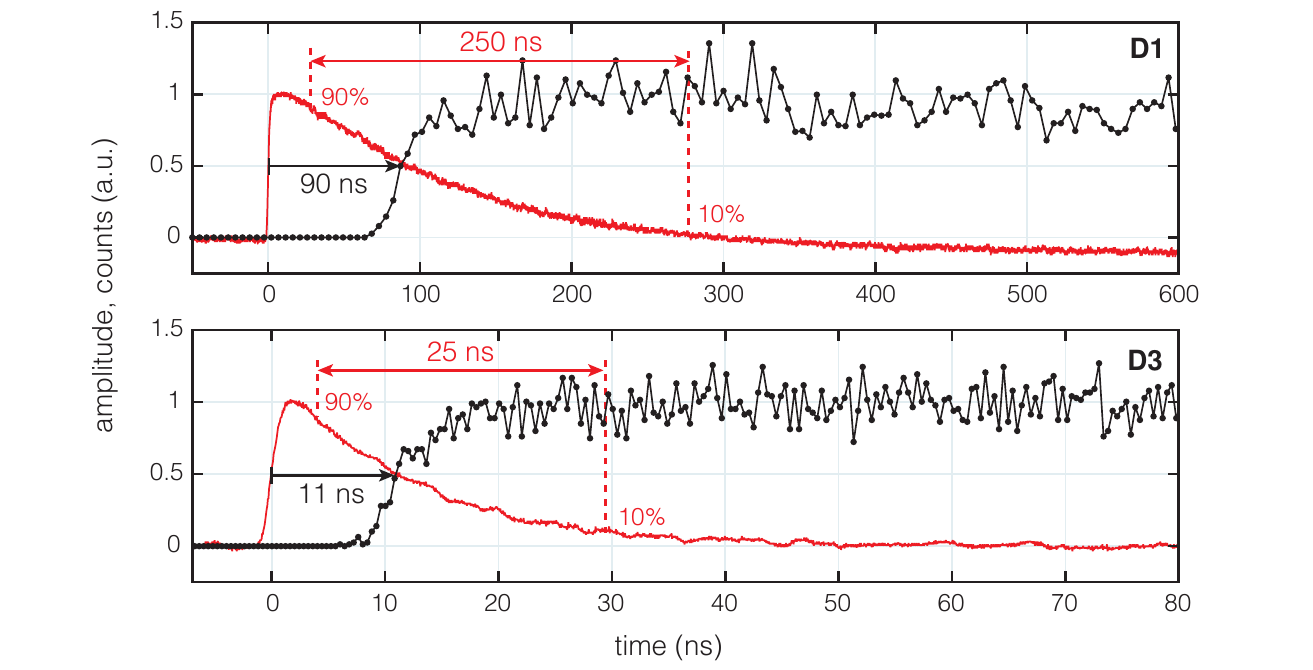}
		\caption{Measurement of detector dead time (black) at 800~mK for large active area (top) and small active area (bottom) devices. At an inter-arrival time of 90~ns for the large device and 11~ns for the small device, there were half as many detection events than for longer inter-arrival times. Typical pulses for the large and small devices, with electrical decay times of 250 and 25~ns, respectively, are shown for comparison in red.}
		\label{fig3}
	\end{figure}
	
	\subsection{Jitter}
	To measure timing jitter, we illuminated the device with a pulsed laser and took a histogram of the delay times between detector pulses and laser pulses. With the 373~nm gain-switched laser, the FWHM of the resulting histogram was $> 140$~ps, but the strong dependence of the jitter on laser power suggested that the contribution of the laser's pulse width was not negligible. We thus used a sub-picosecond mode-locked laser at 1550~nm and a 10~\um~core fiber to measure the jitter. It has been suggested that the intrinsic jitter is wavelength dependent, but some observations have found the jitter to be the same at different wavelengths \cite{Pearlman:2005, Najafi:2015}. If the jitter is dominated by electrical noise in the readout or by geometric effects rather than intrinsic jitter, then the jitter should be independent of wavelength.
	
	For the small active area device (D3), we found a minimum jitter of 62 ps at 800 mK for a 16~\textmu A bias current, with performance deteriorating at lower currents. As the switching current decreases at higher temperatures, this limits the timing properties of the device at higher temperatures. The larger active area device (D1) is expected to have worse jitter for comparable temperatures and bias currents due to additional geometric jitter\cite{Calandri:2016} and the longer rise time of the pulse.
	
	\begin{figure}[ht!]
		\centering\includegraphics[]{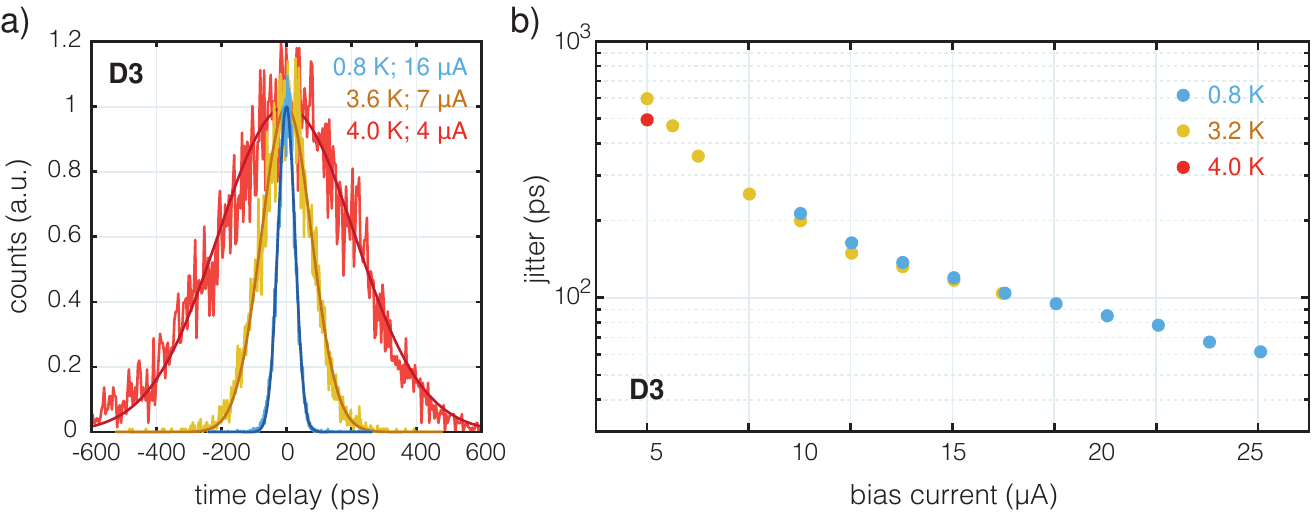}
		\caption{a) Instrument response function near the switching current at 800~mK, 3.6~K, and 4.0~K for the small active area device (D3). The histograms have been fitted with Gaussian distributions. At lower bias currents, the histograms appear slightly skewed. The jitter --- i.e., the full-width at half-maximum of the distribution --- at these temperatures was 62~ps, 182~ps, and 494~ps, respectively. b) Jitter vs. bias current at 800~mK, 3.2~K, and 4.0~K for D3.}
		\label{fig4}
	\end{figure}
	
	\section{Conclusion}
	By adapting SNSPDs for UV wavelengths, we have shown that it is possible to produce detectors with high efficiency ($84 \pm 5\%$ for narrowband designs and $76 \pm 4\%$ for broadband designs), ultra-low dark count rates ($< 10^{-2}$~counts/s system dark counts and $< 10^{-4}$~counts/s isolated dark counts), sub-nanosecond timing resolution, and rejection of out-of-band wavelengths --- all at 4~K. For trapped-ion fluorescence detection, single-pixel SNSPDs can already beat the efficiency and noise of PMTs. With further development to scale this single-pixel technology to larger formats, UV SNSPDs would compete favorably with MCPs, EMCCDs, sCMOS, and other superconducting detectors for photon-starved astronomy applications \cite{France:2016}.
	
	\section*{Acknowledgments}
	The research was carried out at the Jet Propulsion Laboratory, California Institute of Technology, under a contract with the National Aeronautics and Space Administration. The authors would like to thank other members of the Extreme-Performance Ion trap-Cavity System (EPICS) team, including Jungsang Kim, Stephen Crain, Clinton Cahall, and Andre Van Rybach from Duke University, and Christian Arrington, Andrew Hollowell, and Peter Maunz from Sandia National Laboratories. J. P. A. acknowledges support from the NASA Space Technology Research Fellowship program.

	%%%%%%%%%%%%%%%%%%%%%%% References %%%%%%%%%%%%%%%%%%%%%%%%%
	\bibliographystyle{osajnl}
	\bibliography{UVMoSi}
\end{document}